\documentclass[prb,twocolumn,showpacs,preprintnumbers,superscriptaddress]{revtex4-1}
\usepackage[usenames,dvipsnames]{color}
\usepackage{amssymb} 
\usepackage{times}
\usepackage{graphicx}
\usepackage{graphicx}
\usepackage{dcolumn}
\usepackage{datetime}
\usepackage{soul}
\usepackage{subfigure}
\usepackage[bookmarks, colorlinks=true, plainpages = false,citecolor = red, urlcolor = black, 
filecolor = blue]{hyperref}
\usepackage{multirow} 
\usepackage[english]{babel}
\usepackage{graphicx}
\usepackage{amsmath}
 
\usepackage[english]{babel}
\usepackage{graphicx}
\usepackage{amsmath}
 
\begin{document}
\title{Symmetry, Disorder and Transport  Through Altermagnetic  Quantum Dots and Their Antiferromagnetic Twins}
 
\author{George Kirczenow}  

\affiliation{Department of Physics, Simon Fraser
University, Burnaby, British Columbia, Canada V5A 1S6}

\date{\today}

\begin{abstract}\noindent
Altermagnetic crystals resemble antiferromagnets in that they have no macroscopic magnetization, but unlike antiferromagnets they exhibit spin-split band structures. Here the transport properties of altermagnetic quantum dots and  their antiferromagnetic twins are explored theoretically for the first time with the help of Landauer-Buttiker theory, symmetry considerations and tight-binding models. The influence of the symmetries of the quantum dots, their parent crystal lattices, their shapes and edges,  lead arrangements and disorder on the anomalous Hall effect, the spin-Hall effect and spin filtering by the quantum dots are investigated. 
\end{abstract}

 \maketitle

 \section{Introduction}
\label{Intro}

Altermagnets, in common with antiferromagnets, have compensated antiparallel magnetic ordering at the atomic scale and thus have no overall macroscopic magnetization. However, in common with ferromagnets but unlike antiferromagnets,  they have spin-split band structures in reciprocal space.\cite{Bai2024} This difference between altermagnets and antiferromagnets is related to differing symmetry operations  connecting the crystal sublattices of opposite spin in these two classes of materials. \cite{Smejkal2022a,Smejkal2022b,Xiao2024,Chen2024,
Jiang2024,Zhu2025,Che2025,Fender2025} 
The transport phenomena exhibited by altermagnets include the anomalous Hall effect (a transverse voltage induced by an electric current in absence of any applied magnetic field)\cite{Ghimire,Smejkal2020,Naka2020,Reichlova,Feng2022,
Naka2022,Gonzalez2023,Nguyen2023,Hou2023,Fakhredine2023,
Wang2023,Leiviska2024} the spin-Hall effect\cite{Sheng2006,Sinova2015,Mook2020,Naka2019,Ma2021,Gonz2021,MNaka2021,HBai2022,Bose2022,Karube2022,HBai2023,ZHa2024,Bai2024,Cui2023,ZHa2025}  and non-relativistic spin currents. \cite{Naka2019,Ma2021,Gonz2021,MNaka2021,HBai2022,Bose2022,Karube2022,HBai2023,XChen2023,Cui2023,Sun2023,YChe2024,Hodt2024}

These phenomena in {\em macroscopic} altermagnetic crystals are attracting intense theoretical and experimental interest.\cite{Bai2024} However, little or nothing is known regarding potential analogs of the  anomalous Hall effect and the spin-Hall effect or non-relativistic spin currents in ultra small altermagnetic structures, i.e., in quantum dots. Important open questions concern the roles of the quantum dot's geometry, its symmetries, its parent crystal lattice, its shape and structure of its edges, any defects or disorder that may be present, and the configuration of leads that should be attached to the quantum dot for studies of quantum dot  transport. 

\begin{figure}[t]
\centering
\includegraphics[width=0.9\linewidth]{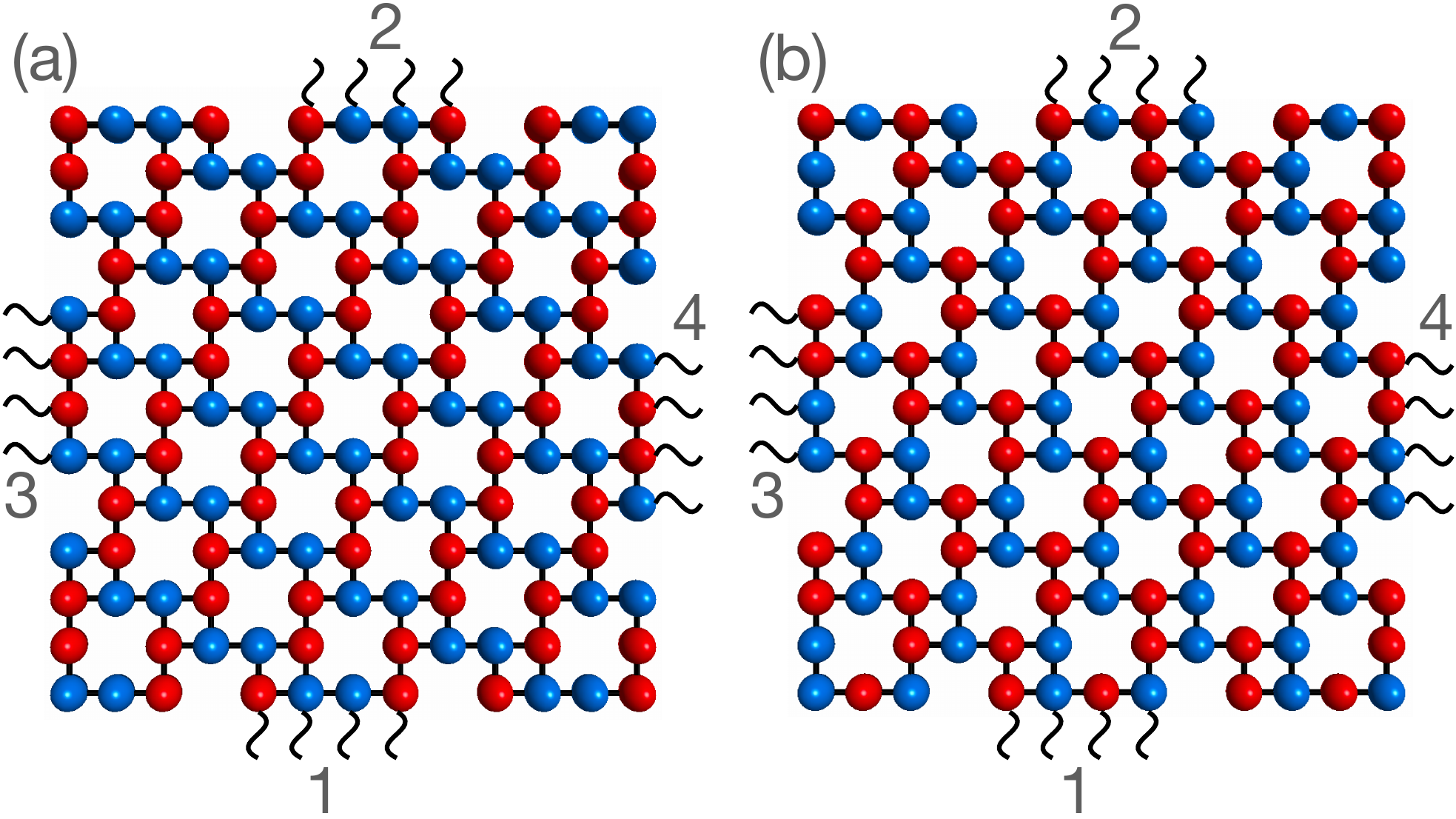}
\caption{(Color online).
(a) An altermagnetic model quantum dot and leads with C$_4$$\mathcal{T}$ symmetry and (b) its antiferromagnetic twin and leads with $\mathcal{IT}$ symmetry. Red (blue) disks are sites with up (down) out of plane local spins.
Wavy lines represent one dimensional  ideal conductors (modeled as tight binding chains) each carrying a spin up and a spin down conducting channel. Four such conductors constitute each electrical lead contacting the quantum dot. In a Hall measurement leads 1 and 2 carry the electric current, leads 3 and 4 carry no  current, and the Hall voltage drop is measured between leads 3 and 4.  The quantum dots (a) and (b) are fragments of the altermagnetic and antiferromagnetic 1/5-depleted square lattice bulk crystals discussed in Ref. \onlinecite{Zhu2025},  respectively. The dot edges are of the armchair type. Images prepared using Macmolplt software.\cite{MacMolPlt}
}
\label{start} 
\end{figure}

This paper initiates the theoretical exploration of these issues by considering tight binding models that have recently been proposed by Zhu {\em et al.}\cite{Zhu2025} and  Che {\em et al.} \cite{Che2025} as vehicles for investigating the fundamental physics
of macroscopic altermagnet crystals in relation to their real space symmetries. Here these models are adapted
to describe nanoscale altermagnetic quantum dots with C$_4$$\mathcal{T}$ symmetry and their antiferromagnetic twins with $\mathcal{IT}$ symmetry. An example of both is shown in Fig.\ref{start}. Here C$_4$, $\mathcal{I}$ and $\mathcal{T}$ are the spatial fourfold rotation, inversion and time reversal operations respectively. Transport through such quantum dots in two terminal and four terminal configurations is modeled here by attaching non-magnetic ideal tight binding leads to the quantum dot. The relevant transport coefficients, namely, the anomalous Hall resistance, the spin-Hall conductance  and the spin filtering efficiency, are calculated within the B\"{u}ttiker-Landauer formalism\cite{BL,FL,ES} after evaluating the Landauer lead-to-lead  electron quantum transmission  probabilities by solving the relevant Lippmann-Schwinger equations that describe electron scattering through the quantum dot. These quantum transport calculations are carried out in the linear response regime for quantum dot geometries based on the
1/5-depleted square lattice,\cite{Zhu2025}  the square-octagon lattice\cite{Zhu2025} and  the Lieb lattice\cite{Che2025}
with arm-chair, zigzag, square or octagon edges and various lead arrangements, conforming to C$_4$$\mathcal{T}$ or $\mathcal{IT}$
symmetries. The effects of disorder in the quantum dots and symmetry breaking lead arrangements are also investigated.

The new results reported here are as follows:

If the complete system of the  altermagnetic  quantum dot together with the four leads attached to it in the Hall geometry is symmetric under the operation C$_4$$\mathcal{T}$ as in Fig.\ref{start}(a) then there is no anomalous Hall effect. I.e., in the 4-lead Hall geometry there is no potential  difference between the two voltage leads when they carry no current while a current flows through the two current leads. This finding is shown here to be a consequence of the  C$_4$$\mathcal{T}$ symmetry. However, the antiferromagnetic twin of this system (that has the same atomic structure but whose distribution of local spins is symmetric under the operation $\mathcal{IT}$ as in Fig.\ref{start}(b)) does exhibit the anomalous Hall effect. Moreover, this antiferromagnetic quantum dot exhibits giant Hall resistances in the tunneling regime. 

Also, the  altermagnetic  quantum dot and its antiferromagnetic twin are predicted here to exhibit a spin-Hall effect provided that the nanostructure consisting of the dot and leads does not have mirror symmetry. In these systems the spin-Hall effect manifests itself as pure spin currents in the Hall voltage leads that carry no net electric currents.

If the Hall leads are removed from these systems and only the current leads are retained in the same geometry as above, then the altermagnetic quantum dot is predicted here to act as an efficient electron spin filter whereas its antiferromagnetic twin is predicted to exhibit no spin filtering due to its $\mathcal{IT}$ symmetry. 

Finally, if all of the symmetries of these systems are broken by introducing disorder into the quantum dot or appropriately altering the lead geometries,
both the altermagnetic quantum dots and their antiferromagnetic twins are predicted to exhibit the anomalous Hall effect, the spin-Hall effect and spin filtering.

The remainder of this paper is organized as follows:
The Hamiltonian and  formalism used to calculate the transport properties of altermagnetic and antiferromagnetic quantum dots are described in Sec. \ref{M}.
Altermagnetism and antiferromagnetism in quantum dots are discussed in Sec. \ref{AandA}.
The anomalous Hall effect, the spin-Hall effect and spin filtering by  altermagnetic and antiferromagnetic 1/5-depleted square lattice quantum dots with armchair edges are discussed in Secs. \ref{AH1}, \ref{SH1} and  
\ref{filter}, respectively. Other  quantum dot edge and lead geometries are discussed in Sec. \ref{other}. Quantum dots with mirror symmetries are discussed in Sec. \ref{mirror}. Effects of disorder in quantum dots on transport are addressed in Sec. \ref{disorderly}. Symmetry breaking leads are discussed in Sec. \ref{break}.
The main findings of this work are summarized in Sec. \ref{S}. 

 \section{Method}
\label{M}
A minimal model Hamiltonian\cite{Zhu2025} for the altermagnetic and antiferromagnetic quantum dots considred in this paper is:
\begin{equation}
H=-t\sum_{\langle ij\rangle} (a_{i\uparrow}^\dagger a_{j\uparrow}+a_{i\downarrow}^\dagger a_{j\downarrow})
-J\sum_i S_i(a_{i\uparrow}^\dagger a_{i\uparrow}-a_{i\downarrow}^\dagger a_{i\downarrow})
\label{Ham}
\end{equation}
where $a_{i\uparrow}^\dagger$ creates a spin up electron on site $i$ of the quantum dot, 
$\langle ij\rangle$ indicates nearest neighbor sites $i$ and $j$, and 
$S_i = S$ for local spin up at site $i$ and $S_i = -S$ for local spin down at site $i$. In this work the atomic chains making up the leads are represented by Hamitonian
\begin{equation}
    H_\text{C}=-t_\text{C}\sum_{\langle ij\rangle} (b_{i\uparrow}^\dagger b_{j\uparrow}+b_{i\downarrow}^\dagger b_{j\downarrow})
\label{chain}
\end{equation}
where $b_{i\uparrow}^\dagger$ creates a spin up electron on site $i$ of the chain. These atomic chains are non-magnetic since there is no energy splitting between spin up and down states in $H_\text{C}$. It is assumed throughout this paper that the spin up and down states traveling through the leads towards the quantum dot are equally populated with electrons and that the temperature is zero Kelvin.

In this work calculation of the transport coefficients of quantum dots described by Eq. \ref{Ham} and \ref{chain} proceeds by solving 
the B\"{u}ttiker equations\cite{BL}
\begin{equation}\label{buttiker}
I_i=\frac{q_e}{h}(N_i\mu_i-R_{ii}\mu_i-\sum_{j\neq i}T_{ij}\mu_j),
\end{equation}
where $I_i$ is the electric current in lead $i$, $\mu_i$ is the electrochemical potential of
lead $i$, $T_{ij}$ is the electron transmission probability from lead $j$ to lead $i$, and $R_{ii}$ is the electron reflection probability from the quantum dot in lead $i$, $N_i$ is the total number of spin up and spin down conducting channels supported by lead $i$ and $q_e$ is the electron charge.

The electron transmission and reflection probabilities $T_{ij}$ and $R_{ii}$ are found by numerically solving the Lippmann-Schwinger equations

\begin{equation}\label{lippmann}
|\psi^m\rangle=|\phi_{\circ}^m\rangle+G_{\circ}(E)W|\psi^m\rangle,
\end{equation}
as is described in Appendix A of Ref. \onlinecite{LSsol}. Here $|\phi_{\circ}^m\rangle$ is an eigenstate of the $m^{th}$ lead that is decoupled from the quantum dot, $G_{\circ}(E)$ is the sum of the Green's functions of the quantum dot and leads when they are decoupled, and $|\psi^m\rangle$ is the scattering eigenstate of the coupled system. $W$ is the coupling between the quantum dot and the ideal leads, i.e.,
\begin{equation}\label{W}
W=-\sum_nt_\text{C}(b_{n\uparrow}^{\dag}a_{n\uparrow}+b_{n\downarrow}^{\dag}a_{n\downarrow}+H.c.)
\end{equation}
where $b_{n\uparrow}^{\dag}$ is the electron creation operator at a lead site attached to the quantum dot, $a_{n\uparrow}$ is the electron annihilation operator at the quantum dot site attached to the corresponding lead, and the hopping amplitude $t_\text{C}$ is assumed for simplivity to be the same as the hopping between the sites of the chains that make up the leads. Having evaluated scattering states $|\psi^m\rangle$, the coefficients $T_{ij}$ that enter B\"{u}ttiker equations (Eq.\ref{buttiker}) are given by 
\begin{equation}\label{Tij}
T_{ij}(E)=\sum_{l,p}|t_{lp}^{ij}|^2 \frac{v_l^i}{v_p^j},
\end{equation}
where $t_{lp}^{ij}$ is the quantum transmission amplitude of an electron transmitted from the $p^{th}$ chain of lead $j$ to the $l^{th}$ chain of lead $i$ at energy $E$ obtained from the scattering states $|\psi^m\rangle$. $v^{i(j)}_{l(p)} = \frac{1}{\hbar} \frac{\partial\epsilon}{\partial k}$ is the electron velocity in the 1-D semi-infinite chain $l(p)$ of lead $i(j)$ at energy $E$, and $\epsilon$ are the energy eigenvalues of the tight-binding Hamiltonian of the semi-infinite ideal chain.

To calculate the Hall resistance $R_\text{H}$ of a quantum dot, following the calculation of electron transmission probabilities $T_{ij}$ at the Fermi energy $E_\text{F}$, the B\"{u}ttiker equations Eq. (\ref{buttiker}) are solved for the potential difference $\Delta V_{3,4}$ between the voltage contacts 3 and 4  subject to the condition of zero electric current in the voltage leads, i.e., $I_3=I_4=0$. Then
\begin{equation}\label{RHall}
R_\text{H}=\frac{\Delta V_{3,4}}{I_{1,2}},
\end{equation} 
where $I_{1,2}$ is the electric current flowing between the current contacts 1 and 2 and; the numbering of the contacts is as in Fig.\ref{start}. Here the potential difference $\Delta V$ and electrochemical potentials $\mu$ appearing in the B\"{u}ttiker equations Eq. (\ref{buttiker}) are related by $\Delta V=\Delta \mu/q_e$. 

Having found the electrochemical potentials $\mu_i$ for which the electric currents in the Hall voltage leads are zero,  $I_3=I_4=0$, the spin-Hall effect is investigated here as follows: For the values of the electrochemical potentials $\mu_i$ for which the electric currents in the Hall voltage leads are zero the spin-resolved version of the B\"{u}ttiker equations
(\ref{buttiker}) is used to evaluate the spin-resolved electric currents $I_{i\uparrow}$ and $I_{i\downarrow}$ in the leads. Then the spin current $I^\text{s}_i$ in voltage lead $i$ is defined as\cite{Sheng2006} 
\begin{equation}\label{spincur}
I^\text{s}_i=\frac{\hbar}{2q_e}(I_{i\uparrow}-I_{i\downarrow})
\end{equation} 
A non-zero spin current $I^\text{s}_i$ in a voltage lead where the total electric current is zero is a manifestation  of the spin-Hall effect.

In this work in order to address spin filtering in 2-terminal geometries  a spin-unpolarized electron flux is considered to enter the quantum dot
through the electron source lead. Then the spin resolved probabilities $T_{\uparrow}$ and 
$T_{\downarrow}$ of spin up and spin down electrons
exiting through the drain lead at the Fermi energy are calculated. $T_{\uparrow}$ and 
$T_{\downarrow}$ are obtained by restricting the sum over $l$ in Eq. (\ref{Tij}) to spin up and spin down states,
respectively, while including both the spin up and spin down states in the sum over $p$. The axis of quantization is perpendicular to the plane of the quantum dot. The spin filtering efficiencies are then defined as

\begin{eqnarray}\label{Fup}
F_{\uparrow}=T_{\uparrow}/(T_{\uparrow}+T_{\downarrow})\\
F_{\downarrow}=T_{\downarrow}/(T_{\uparrow}+T_{\downarrow})
\end{eqnarray}

For the present transport calculations the parameter values in Eq. (\ref{Ham}) are chosen to be $t=1$eV and $JS=0.5$eV as in Ref. \onlinecite{Zhu2025} for sites with local spins. For sites with no local spin $JS=0.$ The parameter $t_\text{C}$ in Eqs. \ref{chain} and \ref{W} has been chosen to be 2.0 eV. This choice ensures that the full spectrum of the quantum dot energy eigenstates falls well within the bandwidth of the tight binding chains that that make up the leads that carry electrons into and out of  the quantum dots.

  \section{Results}
\label{R}
 \subsection{ Altermagnetic and Antiferromagnetic Quantum Dots}
 \label{AandA}
The quantum dots in Fig.\ref{start} (a) and (b) are fragments  of 1/5-depleted square lattice bulk crystals with  C$_4$$\mathcal{T}$ and $\mathcal{IT}$ symmetries, respectively. As has been pointed out in Ref. \onlinecite{Zhu2025}, those  bulk crystals are altermagnetic and antiferromagnetic, respectively. This finding of Ref. \onlinecite{Zhu2025} has been confirmed by band structure calculations carried out for these bulk crystals as a part of the present work, i.e., the bulk crystals with C$_4$$\mathcal{T}$ symmetry are found to exhibit spin split electronic bands and thus are altermagnetic, whereas those with  $\mathcal{IT}$ symmetry do not and are thus antiferromagnetic.  Accordingly, the quantum dots in Fig.\ref{start} (a) and (b) will be referred to as altermagnetic and antiferromagnetic, respectively.

The nanostructures in Fig.\ref{start} (a) and (b) are fully symmetric under the operations  C$_4$$\mathcal{T}$ and $\mathcal{IT}$, respectively. This applies to the quantum dots themselves, including their edges, as well as to the non-magnetic leads. These complete symmetries are decisive in determining the transport properties of these systems, as will be seen below. 

\subsection{Anomalous Hall Effect}
\label{AH1}
\subsubsection{Altermagnetic Quantum Dot}
For the altermagnetic quantum dot in Fig.\ref{start} (a) the numerical calculations described in Section \ref{M} were carried out choosing  leads 1 and 2  to be the current leads and leads 3 and 4 to be the voltage leads that carry no current. The result was $R_\text{H}=0$, i.e., zero Hall resistance defined by Eq. (\ref{RHall}) (and zero Hall voltage) for all values of the electron Fermi energy. 

This absence of  an anomalous Hall effect can be understood as a direct consequence of the C$_4$$\mathcal{T}$ symmetry of the altermagnetic quantum dot and leads in Fig.\ref{start} (a), as will be explained next:

In the absence of magnetic fields, electron lead to lead transmission probabilities obey $T_{ij}=T_{ji}$ for all leads $i$ and $j$ due to time reversal symmetry.\cite{BL} However, the C$_4$$\mathcal{T}$ symmetry of Fig.\ref{start} (a) implies additional symmetries of the transmission and reflection probability matrices $T_{ij}$ and $R_{ii}$ and of  $N_i$ the total number of conducting channels supported by each lead.  Namely, 

\begin{equation}\label{NN}
N_{1}=N_{2}=N_{3}=N_{4}\equiv N~~~~~~~~~~~~~~~~~~~~~~~~~~~~~~~~~~~~~~~~~~~~~~~~~~~
\end{equation}
\begin{equation}\label{RR}
R_{11}=R_{22}=R_{33}=R_{44}\equiv R~~~~~~~~~~~~~~~~~~~~~~~~~~~~~~~~~~~~~~~~~~~~~~
\end{equation}
\begin{equation}\label{SS}
T_{12}=T_{21}=T_{34}=T_{43}\equiv Z~~~~~~~~~~~~~~~~~~~~~~~~~~~~~~~~~~~~~~~~~~~~~~~~~
\end{equation}
\begin{equation}\label{TT}
T_{13}=T_{31}=T_{24}=T_{42}=T_{14}=T_{41}=T_{23}=T_{32}\equiv T~~~~~
\end{equation}
Setting $I_3=I_4=0$ for the currents in the voltage leads and solving the B\"{u}ttiker equations (\ref{buttiker})\cite{BL} subject to the symmetry constraints
(\ref{NN}-\ref{TT}) yields $\mu_3=\mu_4=T(\mu_1+\mu_2)/(N-R-Z)$. Applying  the charge coservation condition $N=R+Z+2T$, this reduces to 
\begin{equation}\label{Hallchem}
\mu_3=\mu_4=(\mu_1+\mu_2)/2.
\end{equation}
This means that there is no potential difference between the two voltage leads 3 and 4 and hence no anomalous Hall effect.  

This can be understood intuitively as following from the symmetries between leads 3 and 4 expressed by Eq. (\ref{NN}-\ref{TT}). For example, the condition (\ref{TT}) that $T_{31}=T_{41}$ means that electrons entering the dot from lead 1 are scattered equally into voltage leads 3 and 4 so that cancelation of the associated net currents flowing in leads 3 and 4 can be accomplished by applying equal voltages to leads 3 and 4. 

Thus the C$_4$$\mathcal{T}$ symmetry that results in altermagnetism in a 1/5-depleted square lattice bulk crystal  \cite{Zhu2025} suppresses the anomalous Hall effect in an altermagnetic quantum dot that, together with its leads, obeys C$_4$$\mathcal{T}$ symmetry. It is worth noting that since the structure in Fig.\ref{start} (a) does not have mirror symmetry between the voltage contacts 3 and 4, it is clearly the C$_4$$\mathcal{T}$ symmetry together  with time reversal symmetry $T_{ij}=T_{ji}$\cite{BL} that is responsible for the absence of the anomalous Hall effect. Since this result is purely a consequence of symmetry it holds for all values of the Fermi energy.

\subsubsection{Antiferromagnetic Quantum Dot}

\begin{figure}[t]
\centering
\includegraphics[width=0.9\linewidth]{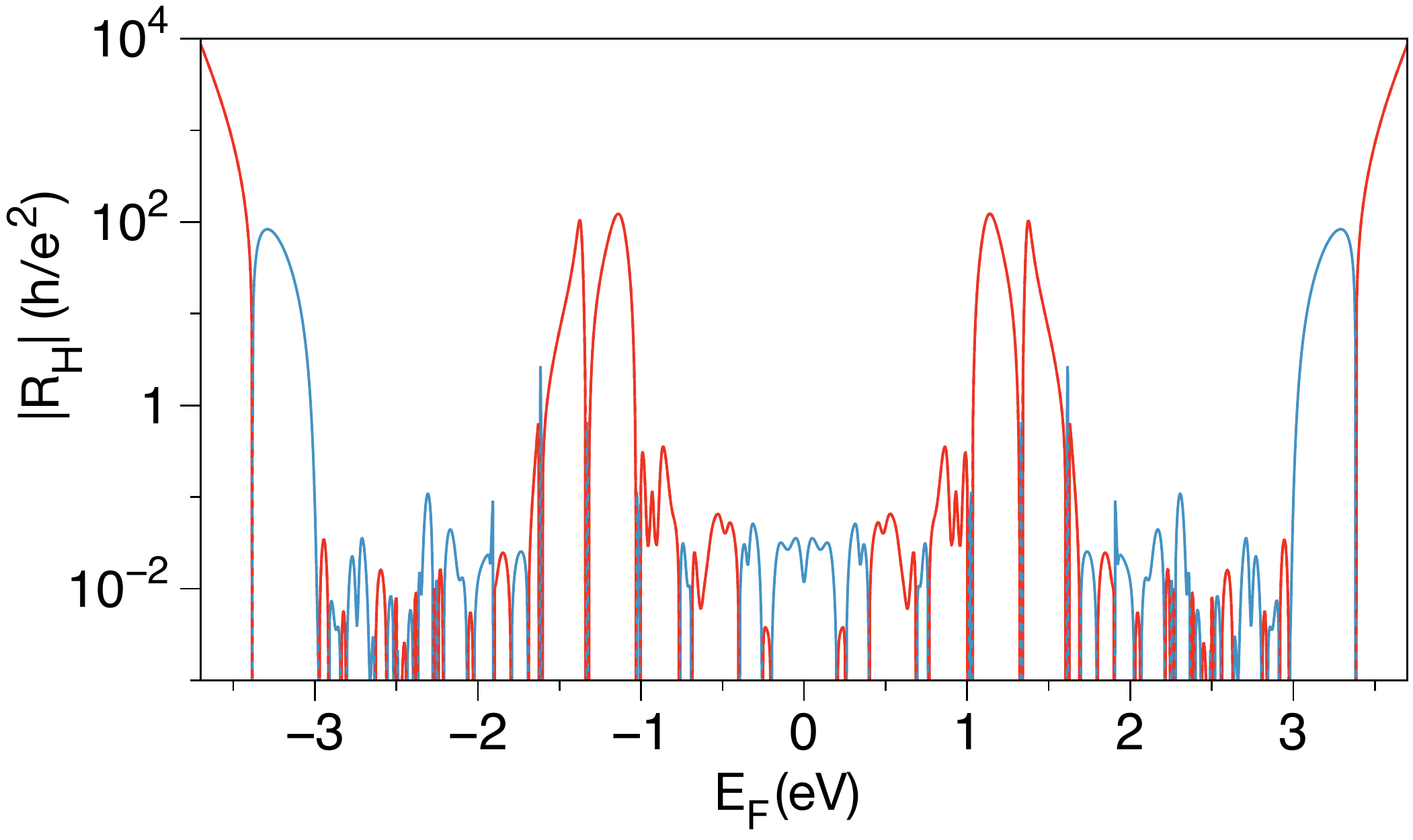}
\caption{(Color online).
Absolute values of the Hall resistance vs. electron Fermi energy of the antiferromagnetic nanostructure in Fig.\ref{start} (b). Positive (negative) values of $R_\text{H}$ are shown in red (blue).
}
\label{AnomHall} 
\end{figure}

The antiferromagnetic quantum dot and leads in Fig.\ref{start} (b) have $\mathcal{IT}$ symmetry, a lower symmetry than that of the  C$_4$$\mathcal{T}$-symmetric altermagnetic structure in Fig.\ref{start} (a). Thus not all of the symmetries expressed by Eq. (\ref{RR}-\ref{TT}) apply. For example,  $T_{31}=T_{41}$ does {\em not} hold for the antiferromagnetic quantum dot and leads. 
Consequently an electron entering the dot from current lead 1 is scattered with unequal probabilities into voltage leads 3 and 4. Thus in order for the net electric currents in the voltage leads to be zero ( $I_3=I_4=0$)  unequal potentials need to be applied to the two voltage leads. Thus the antiferromagnetic quantum dot in Fig.\ref{start} (b) exhibits an anomalous
Hall effect. For this system the  Hall resistance $R_\text{H}$ given by Eq. (\ref{RHall}) has been computed as described in Section \ref{M} and the result is shown in Fig.\ref{AnomHall} where positive (negative) values of $R_\text{H}$ are shown in red (blue).

As is seen in Fig.\ref{AnomHall}, the sign of the Hall resistance depends on the value of the Fermi energy and $R_\text{H}$ exhibits many resonances
that are due in part to the influence of the discrete spectrum of the dot on transport.  Giant values of  $R_\text{H}$ are seen in Fig.\ref{AnomHall} for Fermi energies  $E_\text{F}$ such that $|E_\text{F}| > 3$eV and 1eV$< |E_\text{F}| \lesssim 1.5$eV. This can be interpreted as follows: 

In these energy ranges the density of states for the bulk antiferromagnet of which the quantum dot in Fig.\ref{start} (b) is a fragment, vanishes. Thus transport through the quantum dot at these energies occurs by quantum tunneling. Now consider an electron entering the dot through current lead 1. To exit the dot through current lead 2 and thus to contribute to the current through the dot the electron has to tunnel further than it has to in order to reach voltage lead 3 or 4. For this reason the electric current through the dot decreases more rapidly  than the Hall voltage as the Fermi energy moves more deeply into the tunneling regime. Thus the Hall resistance $R_\text{H}=\frac{\Delta V_{3,4}}{I_{1,2}}$ grows exponentially. For example, as the Fermi energy descends to progressively lower values below -3eV in Fig.\ref{AnomHall} the absolute value of the anomalous Hall resistance is seen to increase by more than 6 orders of magnitude.  

\subsection{Spin-Hall Effect}
\label{SH1}

\begin{figure}[t]
\centering
\includegraphics[width=0.8\linewidth]{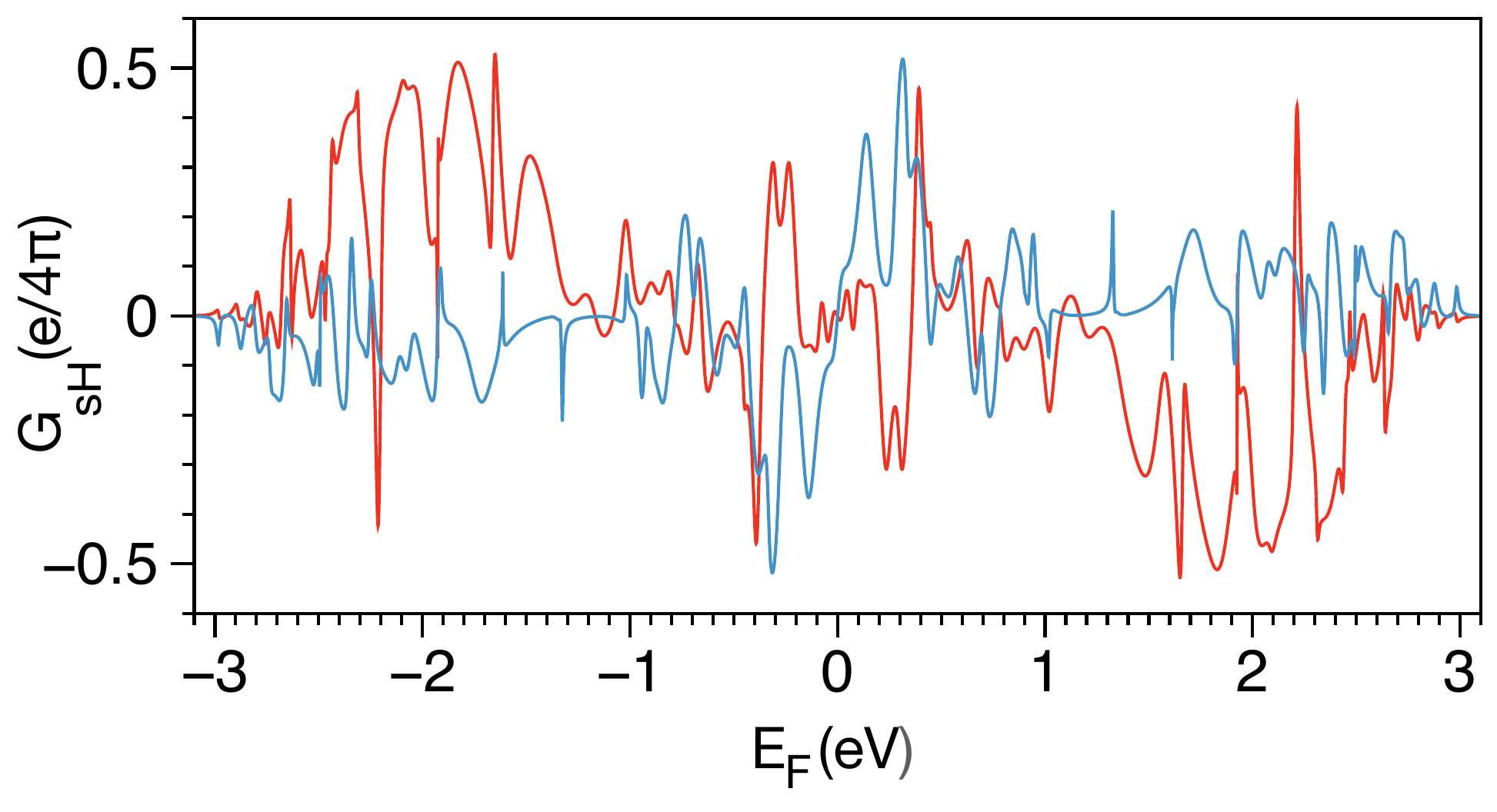}
\caption{(Color online).
Spin-Hall conductance vs. electron Fermi energy of the altermagnetic  (antiferromagnetic) nanostructure in Fig.\ref{start} (a) (Fig.\ref{start} (b)) shown in red (blue).
}
\label{spinHall} 
\end{figure}

For the altermagnetic quantum dot of Fig.\ref{start}(a), for values of the lead electrochemical potentials $\mu_i$ that satisfy Eq. (\ref{Hallchem}) so that the electric currents in the voltage leads 3 and 4 vanish  ($I_3=I_4=0$), the spin-resolved electric currents $I_{i\uparrow}$ and $I_{i\downarrow}$ in the voltage leads have been calculated as described in Section \ref{M}. It was found that $I_{3\uparrow}=-I_{3\downarrow}$ and $I_{4\uparrow}=-I_{4\downarrow}$ in linear response for all values of the electron Fermi energy. This means that there is a net spin current $I^\text{s}_i$ defined by Eq. (\ref{spincur}) in each voltage lead although these leads carry no electric current.  Thus the altermagnetic nanostructure in  Fig.\ref{start} (a) exhibits a spin-Hall effect although it does not support an anomalous Hall effect.
The spin currents in voltage leads 3 and 4 of Fig.\ref{start} (a) are found to be equal and to travel in the same direction, but the direction depends on the Fermi energy.  Fig. \ref{spinHall} shows in red the computed spin-Hall conductance $G_\text{sH}=I^\text{s}/\Delta V$ where
$I^\text{s} $ is the spin current in the leads 3 and 4  of Fig.\ref{start} (a) and $\Delta V$ is the potential difference between leads 1 and 2. 

For the antiferromagnetic dot and leads shown in Fig.\ref{start}(b),
the electrochemical potentials $\mu_i$ for which the electric currents vanish 
in the voltage leads  ($I_3=I_4=0$) were calculated by solving the
B\"{u}ttiker equations (\ref{buttiker}). The spin-resolved electric currents $I_{i\uparrow}$ and $I_{i\downarrow}$ in the voltage leads were then calculated as described in Section \ref{M}. It was found  that $I_{3\uparrow}=-I_{3\downarrow}$ and $I_{4\uparrow}=-I_{4\downarrow}$.  Thus (as for the altermagnetic dot) there is a net spin current $I^\text{s}_i$ defined by Eq. (\ref{spincur}) in each voltage lead although these leads carry no electric current.  The spin currents  $I^\text{s}_i$ in the two voltage leads are equal in 
magnitude but flow in opposite directions which depend on the Fermi energy.
The  spin-Hall conductance $G_\text{sH}=I^\text{s}/\Delta V$ for a voltage lead  of Fig.\ref{start} (b) is shown in blue in Fig. \ref{spinHall}.

Thus both the altermagnetic and antiferromagnetic quantum dots exhibit the spin-Hall effect and both can in principle be used as sources of pure spin currents in Hall voltage leads that carry no electric current.

\subsection{Spin Filtering}
\label{filter}

\begin{figure}[t]
\centering
\includegraphics[width=1.0\linewidth]{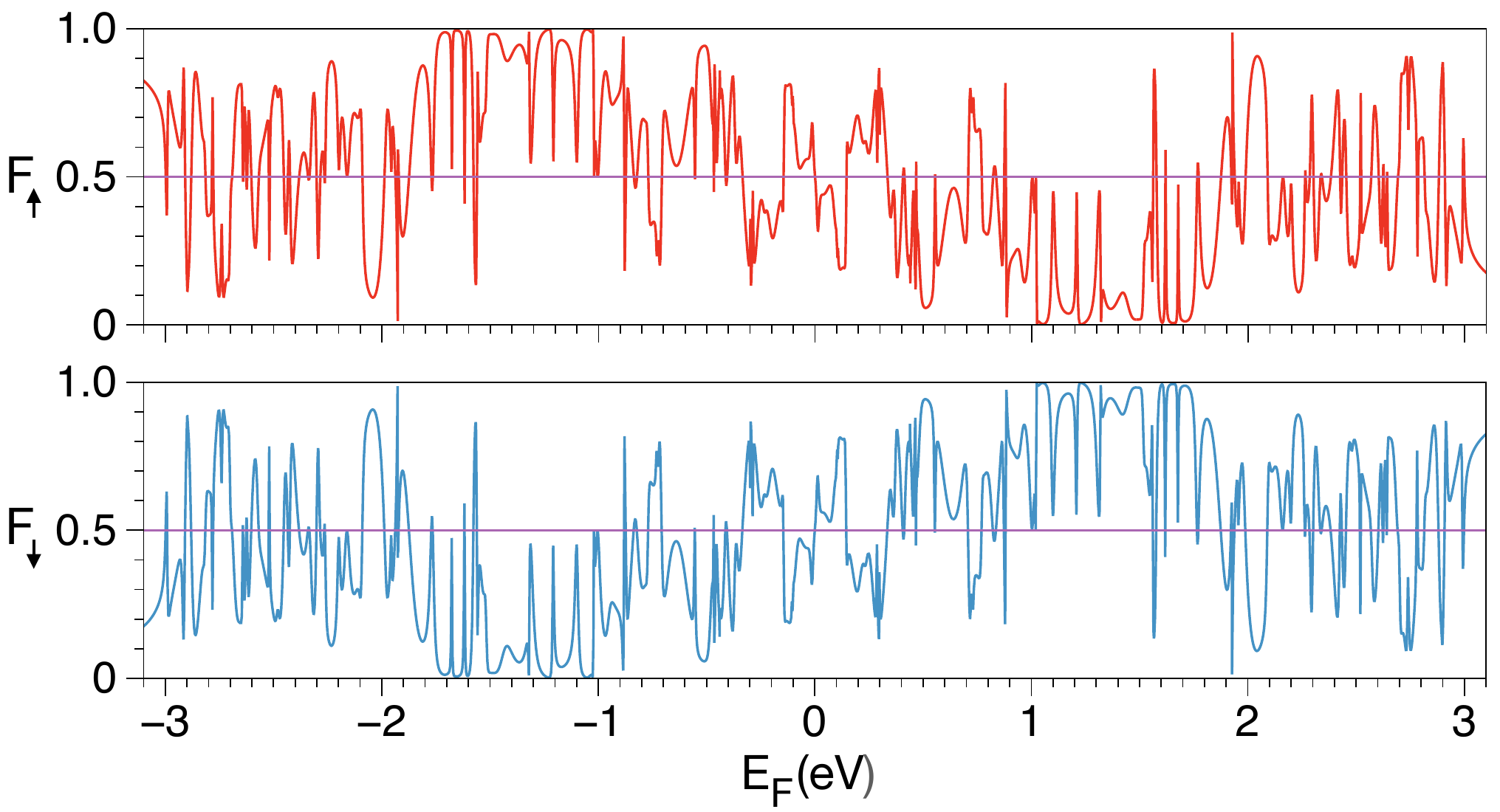}
\caption{(Color online).
Spin filtering efficiencies $F_{\uparrow}$ (red) and 
$F_{\downarrow}$ (blue) vs. electron Fermi energy for the altermagnetic quantum dot in Fig.\ref{start}(a). $F_{\uparrow}$ and 
$F_{\downarrow}$ for the antiferromagnetic dot in Fig.\ref{start} (b)are shown in mauve.}
\label{spinFltr} 
\end{figure}

In order to study spin filtering by altermagnetic and antiferromagnetic quantum dots in two-terminal geometries, the current leads 1 and 2 in Fig.\ref{start} (a) and (b) are retained while leads 3 and 4 are removed.
Both spin up and spin down electrons are assumed to flow towards the dot equally through lead 1. The spin resolved probabilities $T_{\uparrow}$ and 
$T_{\downarrow}$ of spin up and spin down electrons
exiting through lead 2 at the Fermi energy and the spin filtering efficiencies  $F_{\uparrow}$ and 
$F_{\downarrow}$ have been computed numerically as described at the end of Section \ref{M}.

The results for the altermagnetic quantum dot are shown in red and blue in Fig. \ref{spinFltr}. The altermagnetic quantum dot is found to exhibit nearly perfect spin filtering, i.e, $F_{\uparrow} \approx 1$ or 
$F_{\downarrow}\approx 1$ in some ranges of the Fermi energy. By contrast, the antiferromagnetic dot displays no spin filtering at all. For it $F_{\uparrow}$ and 
$F_{\downarrow}$ (shown in mauve) are both 0.5 at all energies.  

This difference between the altermagnetic quantum dot and its antiferromagnetic twin can be understood as follows: 

Application of the symmetry operation $\mathcal{IT}$ of the antiferromagnetic dot transforms lead 2 into
lead 1 and spin up into spin down (and vice versa). It follows that  
\begin{equation}\label{IT}
T_{2\uparrow1\uparrow}=T_{1\downarrow2\downarrow},
\end{equation}
i.e., the transmission probability of a spin up electron from lead 1 to lead 2 equals
 the transmission probability of a spin down electron from lead 2 to lead 1. Noting that the Hamiltonian in Section \ref{M} is diagonal in spin, time reversal symmetry applied to only the spatial part of the wave function  implies that
\begin{equation}\label{TrevSpatial}
T_{1\downarrow2\downarrow}=T_{2\downarrow1\downarrow}
\end{equation}
Eq. (\ref{IT}) and (\ref{TrevSpatial}) 
together yield 
$T_{2\uparrow1\uparrow}=T_{2\downarrow1\downarrow}$. I.e., The transmission probability of a spin up electron from lead 1 to lead 2 equals that for a spin down electron. Since the Hamiltonian in Section \ref{M} is diagonal in spin this means that there cannot be any spin filtering. 

This argument does not hold for the altermagnetic quantum dot  since for it the symmetry operation that connects leads 1 and 2 is not $\mathcal{IT}$ but
(C$_4$$\mathcal{T})^2 = \mathcal{I}$ which does not flip the electron's spin. Therefore the symmetry of the altermagnetic quantum dot does not require the transmission probability of a spin up electron from lead 1 to lead 2 to equal that for a spin down electron, so spin filtering is allowed as is seen in Fig. \ref{spinFltr}. 

\subsection{Other Symmetric Quantum Dot and Lead Arrangements }
\label{other}

\begin{figure}[t]
\centering
\includegraphics[width=0.9\linewidth]{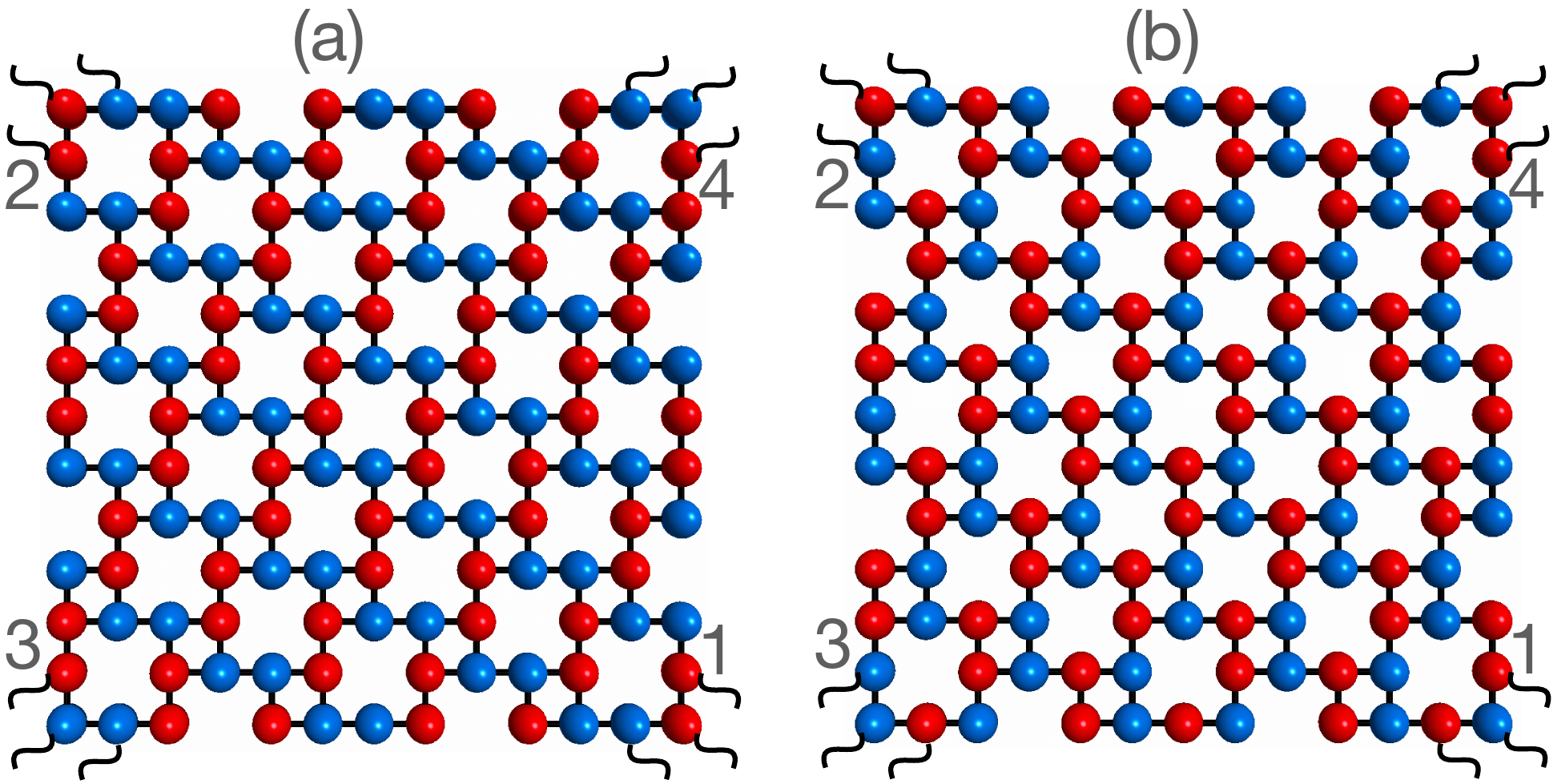}
\caption{(Color online).
Quantum dots (a) and (b) are fragments of the altermagnetic and antiferromagnetic 1/5-depleted square lattice bulk crystals respectively, as in Fig.\ref{start} but with leads attached to the corners of the dots, while preserving the overall C$_4$$\mathcal{T}$ symmetry and $\mathcal{IT}$ symmetry, respectively. Notation as in Fig.\ref{start}.  
}
\label{cornerlds} 
\end{figure}

\begin{figure}[t]
\centering
\includegraphics[width=0.9\linewidth]{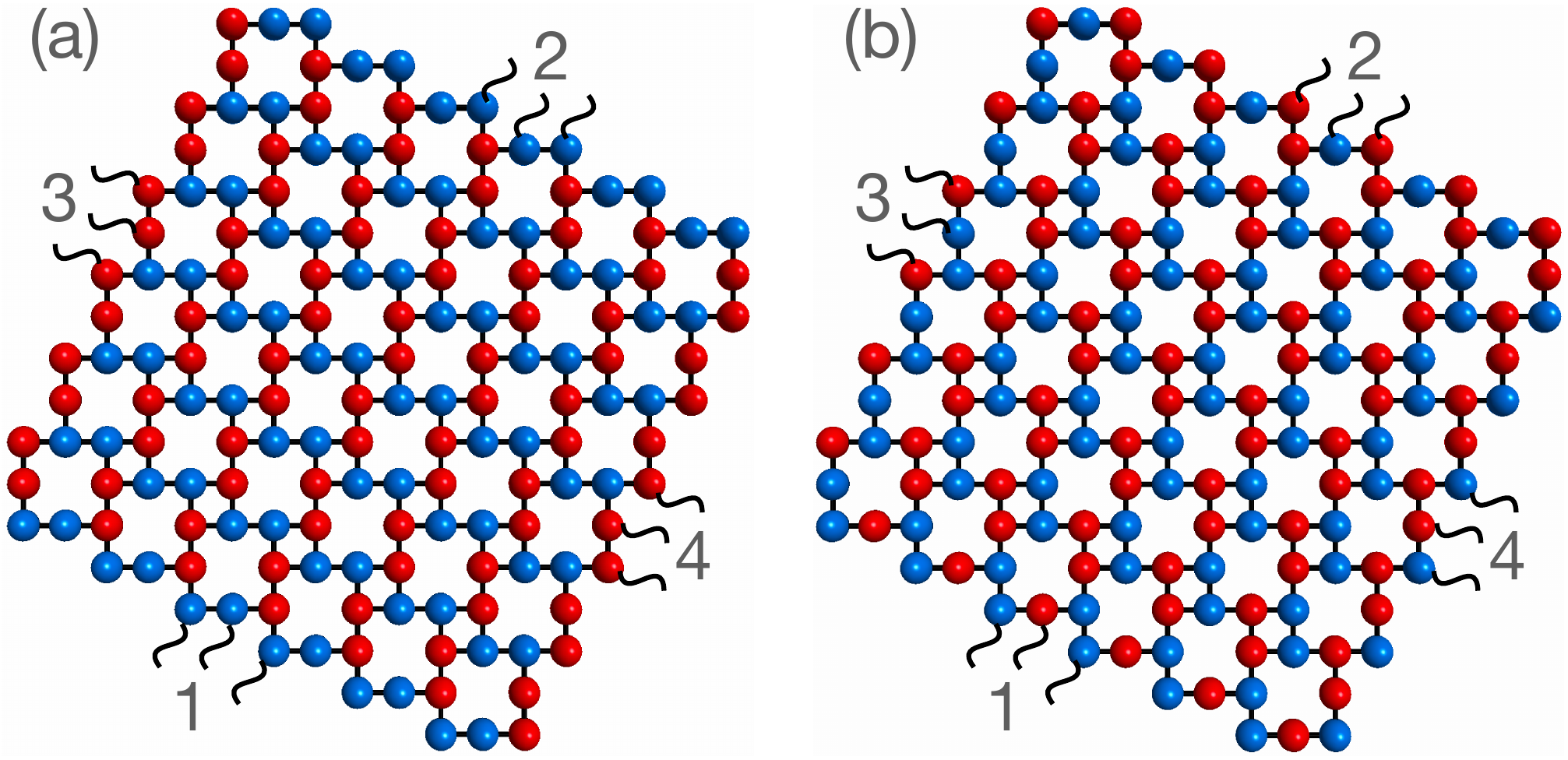}
\caption{(Color online).
Quantum dots (a) and (b) are fragments of the altermagnetic and antiferromagnetic 1/5-depleted square lattice bulk crystals respectively, as in Fig.\ref{start} but with zigzag edges, still preserving the overall C$_4$$\mathcal{T}$ symmetry and $\mathcal{IT}$ symmetry, respectively. Notation as in Fig.\ref{start}.  
}
\label{zigzag} 
\end{figure}
Fig.\ref{cornerlds} (a) and (b) show the same altermagnetic and antiferromagnetic quantum dots as those of Fig.\ref{start} (a) and (b) respectively, but with leads attached to the corners of the dots. The dots with these lead geometries have the same the overall C$_4$$\mathcal{T}$ and $\mathcal{IT}$ symmetries, respectively, as Fig.\ref{start} (a) and (b). Fig.\ref{zigzag} (a) and (b) show dots with zigzag edges. They are different fragments of the altermagnetic and antiferromagnetic 1/5-depleted square lattice bulk crystals, respectively, than those of Fig.\ref{start} but these dots and leads also have the same the overall C$_4$$\mathcal{T}$ and $\mathcal{IT}$ symmetries, respectively.

The present numerical transport calculations for the altermagnetic dot/lead arrangements in Fig.\ref{cornerlds} (a) and Fig.\ref{zigzag} (a) show that they do not exhibit an anomalous Hall effect but do exhibit a spin-Hall effect, and also spin filtering in a two terminal arrangement. The antiferromagnetic structures in 
Fig.\ref{cornerlds} (b) and Fig.\ref{zigzag} (b) are found to exhibit an anomalous Hall effect with giant Hall resistance, a spin-Hall effect, and no two-terminal spin filtering. This behavior is similar to that described above for the corresponding nanostructures in Fig.\ref{start}. This finding suggests that these transport effects are governed by the C$_4$$\mathcal{T}$ and $\mathcal{IT}$ symmetries of the dots and leads, and that provided that these symmetries are obeyed, other details such as the specific shapes of the dots and lead arrangements are not crucial. However, the systems considered above do not have any mirror symmetries. These and their influence on transport  will be addressed next.

\subsection{Quantum Dots with Mirror Symmetries}
\label{mirror}

\begin{figure}[t]
\centering
\includegraphics[width=0.9\linewidth]{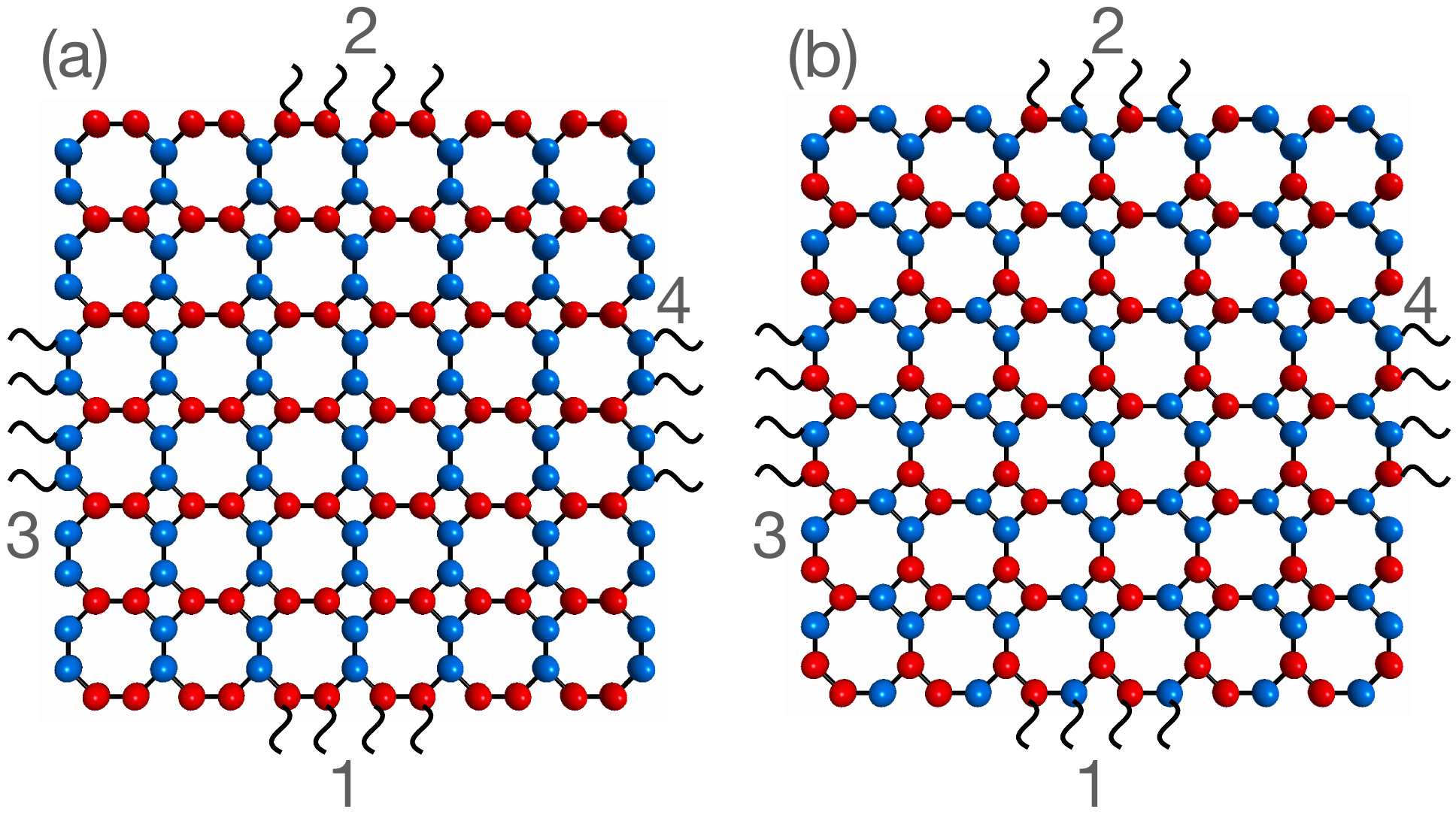}
\caption{(Color online).
Quantum dots (a) and (b) are fragments of altermagnetic and antiferromagnetic square-octagon lattice\cite{Zhu2025} bulk crystals, respectively. These dots and their leads have overall C$_4$$\mathcal{T}$ and $\mathcal{IT}$ symmetries, respectively. Dot edges are of the octagon type. Notation as in Fig.\ref{start}.  
}
\label{sqoct} 
\end{figure}

\begin{figure}[t]
\centering
\includegraphics[width=0.9\linewidth]{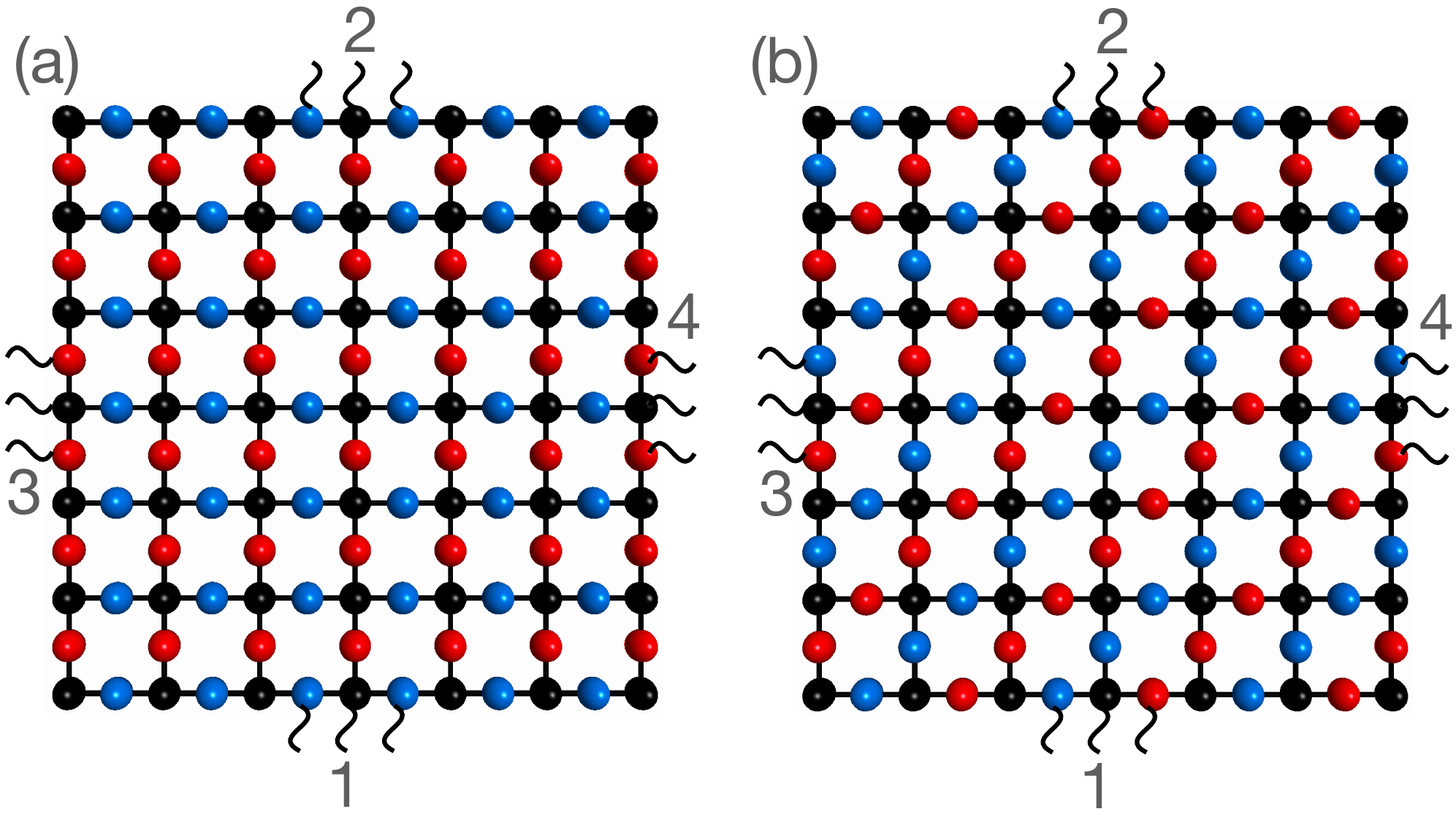}
\caption{(Color online).
Quantum dots (a) and (b) are fragments of altermagnetic and antiferromagnetic Lieb lattice\cite{Che2025} bulk crystals respectively. These dots and their leads have overall C$_4$$\mathcal{T}$ symmetry and $\mathcal{IT}$ symmetries, respectively. Red, blue and black disks represent sites with up, down and no local spins. Dot edges are of square type. Notation as in Fig.\ref{start}.  
}
\label{lieb} 
\end{figure}

Fig.\ref{sqoct} (a) and Fig.\ref{lieb} (a) show altermagnetic quantum dots
that are fragments of  square-octagon lattice\cite{Zhu2025} and Lieb lattice\cite{Che2025} bulk crystals, respectively. These dots and their leads have the C$_4$$\mathcal{T}$ symmetry but in addition they are mirror-symmetric. Their axes of reflection bisect them both horizontally and vertically. Like the other altermagnetic quantum dots discussed above, the present study  has found these systems to not exhibit an anomalous Hall effect, but to support 2-terminal spin filtering if leads 3 and 4 are removed. However, unlike the other altermagnetic quantum dots discussed above these systems do not exhibit a spin-Hall effect. If the mirror symmetries in Fig.\ref{sqoct} (a) and Fig.\ref{lieb} (a) are broken while retaining  the C$_4$$\mathcal{T}$ symmetry by shifting all of the leads clockwise or counter-clockwise by the same amount along the dot's edges, the spin-Hall effect is recovered while   the  anomalous Hall effect remains absent and 2-terminal spin filtering is retained. Thus it is reasonable to conclude that the mirror symmetries are responsible for the suppression of the spin-Hall effect in these systems. This conclusion is consistent with the finding of the present numerical calculations that the antiferromagnetic twin quantum dots/leads in Fig.\ref{sqoct} (b) and Fig.\ref{lieb} (b) (that are symmetric under $\mathcal{IT}$ but have no mirror symmetries) exhibit the spin-Hall effect and anomalous Hall effect but not two terminal spin filtering,  just like the antiferromagnetic quantum dots in Figs.\ref{start}, \ref{cornerlds} and \ref{zigzag} discussed above. 

\begin{figure}[t]
\centering
\includegraphics[width=0.9\linewidth]{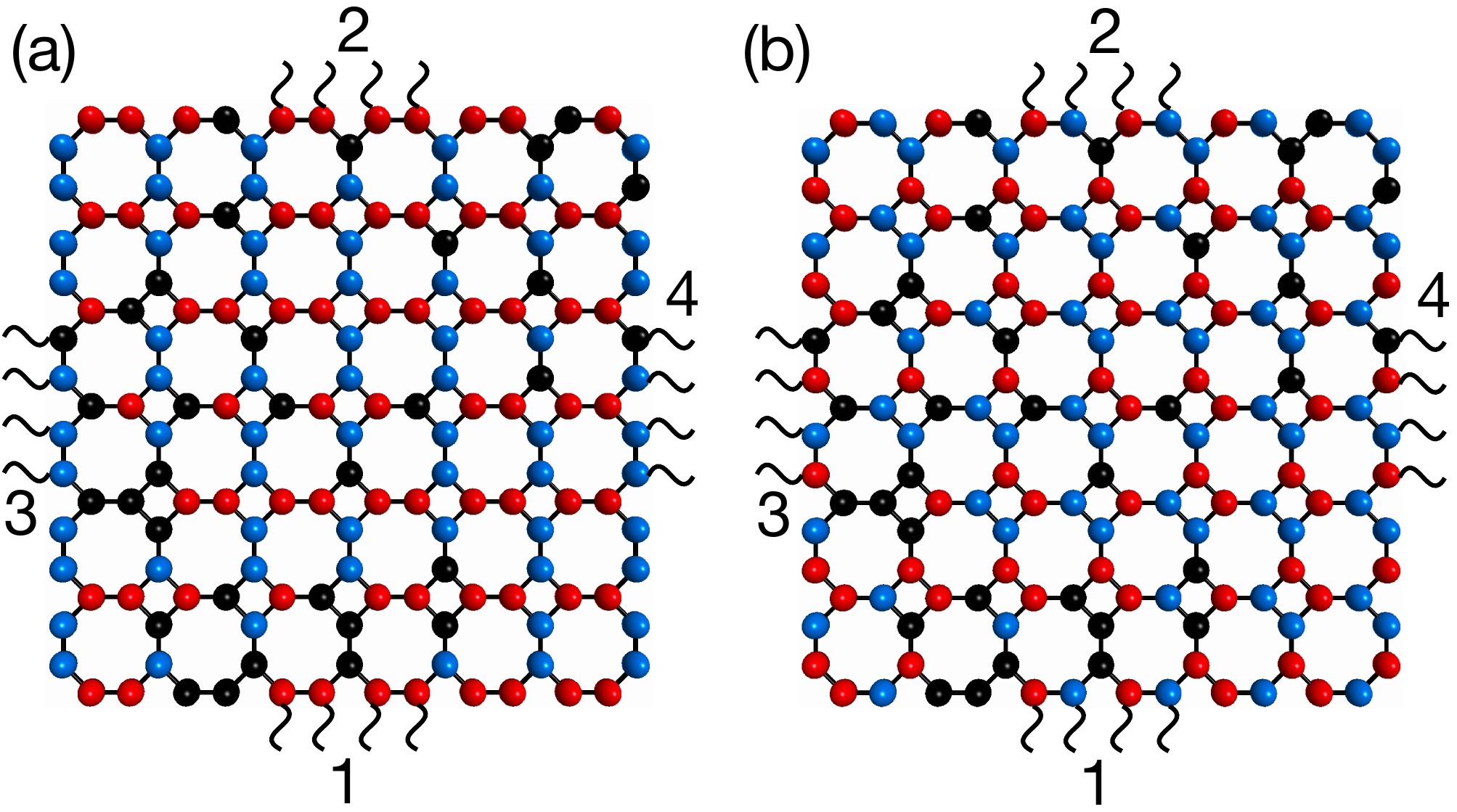}
\caption{(Color online).
Quantum dots (a) and (b) are defective fragments of altermagnetic and antiferromagnetic square-octagon lattice\cite{Zhu2025} bulk crystals, respectively. Their respective C$_4$$\mathcal{T}$ and $\mathcal{IT}$ symmetries and their mirror symmetries are broken by the random substitution of local spins by no-spin defects. Red, blue and black disks represent sites with up, down and no local spins.Notation as in Fig.\ref{start}.  
}
\label{dis} 
\end{figure}

\subsection{Disordered Altermagnetic and Antiferromagnetic Quantum Dots}
\label{disorderly}
The results presented above are for ideal quantum dots and leads that are perfectly  symmetric under  C$_4$$\mathcal{T}$ or $\mathcal{IT}$ operations and that in some cases also have mirror symmetries. However, it is also of interest to know how the transport phenomena are affected if symmetry-breaking disorder is present. To this end the effects of disorder modeled by replacing up and down local local spins at random sites by no-spin sites (where $JS=0$) are examined below. 
Representative examples where $\sim 20\%$ of the local up (red) and down (blue) spins of altermagnetic and antiferromagnetic fragments of square-octagon lattices\cite{Zhu2025} are randomly replaced by no spin (black) sites are shown in Fig.\ref{dis} (a) and (b) respectively. 

\subsubsection{Anomalous Hall Effect }

\begin{figure}[t]
\centering
\includegraphics[width=0.9\linewidth]{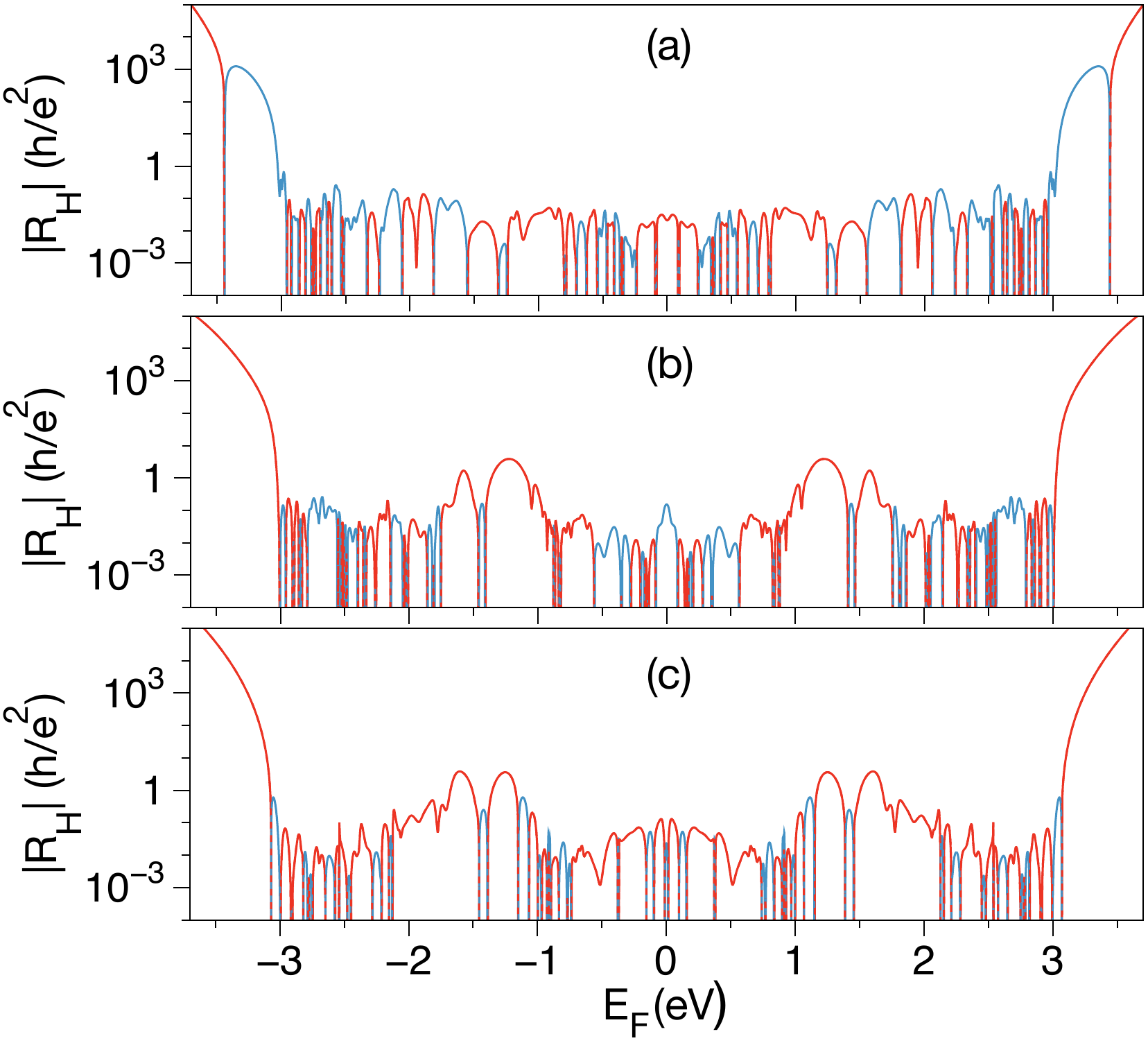}
\caption{(Color online).
Frames (a) and (b) show the absolute values of the Hall resistance vs. electron Fermi energy of the disordered altermagnetic  and antiferromagnetic nanostructures in Fig.\ref{dis} (a) and (b), respectively. Positive (negative) values of $R_\text{H}$ are shown in red (blue). Frame (c) shows for comparison the absolute values of the Hall resistance of the pristine antiferromagnetic nanostructure shown in Fig.\ref{sqoct} (b).
}
\label{Halldis} 
\end{figure}

The calculated anomalous Hall resistances for the disordered altermagnetic and antiferromagnetic quantum dots in Fig.\ref{dis} (a) and (b) are shown in Fig.\ref{Halldis} (a) and (b), respectively.

As was discussed above, the anomalous Hall resistance for the ideal altermagnetic dot in
Fig.\ref{sqoct} (a) is zero for all values of the Fermi energy due to C$_4$$\mathcal{T}$ symmetry. By contrast, as can be seen in Fig.\ref{Halldis} (a), $R_\text{H}$ for the disordered  altermagnetic dot in Fig.\ref{dis} (a) is comparable in magnitude to that of its antiferrormagnetic twin shown in  Fig.\ref{Halldis} (b) except around $|E_\text{F}|=1.5$eV where the bulk antiferromagnetic crystal has a band gap. Both the  disordered  altermagnetic dot and its antiferrormagnetic twin exhibit giant anomalous Hall resistance for $|E_\text{F}| > 3$eV where conduction occurs by tunneling. However, the disorder affects the anomalous Hall resistance of the antiferrormagnetic dot less drastically, as can be seen by comparing Fig.\ref{Halldis} (b) and (c), the disordered and ideal cases, respectively. This is because the C$_4$$\mathcal{T}$ symmetry of the ideal altermagnetic dot is already broken in the ideal antiferrormagnetic dot so that the effects of additional symmetry breaking by disorder are milder: There is some broadening of the 
prominent anomalous Hall resistance peaks around $|E_\text{F}|=1.5$eV
in the disordered case (Fig.\ref{Halldis} (b)) relative to the ideal case (Fig.\ref{Halldis} (c)) as well as reorganization of the fine details of resonance structures.

\begin{figure}[t]
\centering
\includegraphics[width=0.8\linewidth]{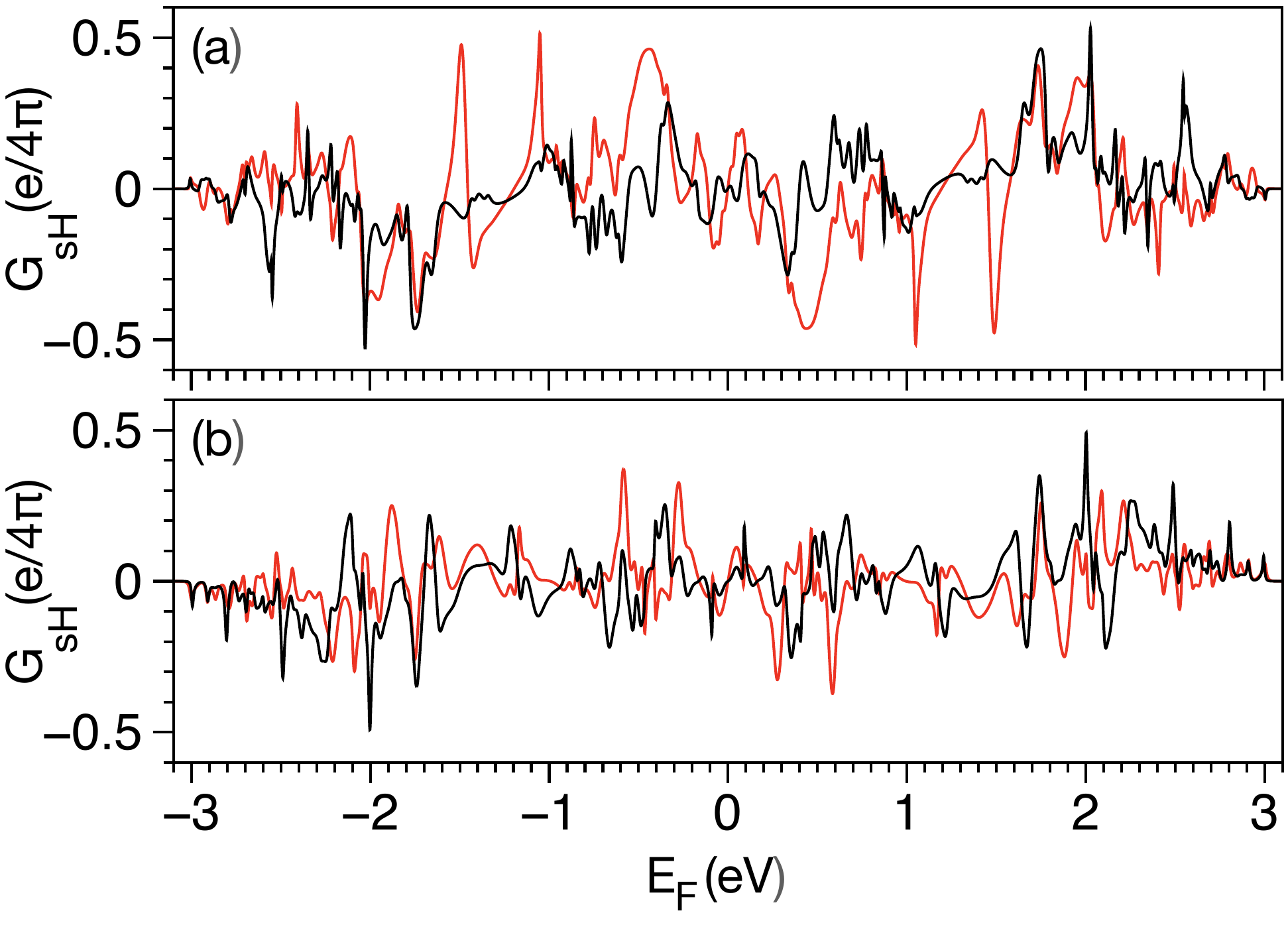}
\caption{(Color online).
Frames (a) and (b) show the spin-Hall conductances vs. electron Fermi energy of the disordered antiferromagnetic and altermagnetic  nanostructures
of Fig.\ref{dis} (b) and (a), respectively.  Values for spin-Hall conductances in voltage leads 3 and 4 are shown in red and black, respectively.
}
\label{spinHalldis} 
\end{figure}

\subsubsection{Spin-Hall Effect }

As was noted above, for the ideal altermagnetic dot in Fig.\ref{sqoct} (a) the spin-Hall conductance is zero due to that structure's mirror symmetry while for its antiferromagnetic twin in Fig.\ref{sqoct} (b) there is no mirror symmetry and consequently the spin-Hall effect is present. The calculated spin-Hall conductances for the disordered antiferromagnetic and altermagnetic  nanostructures
of Fig.\ref{dis} (b) and (a) are shown in Fig. \ref{spinHalldis} (a) and (b),
respectively. The disorder breaks the mirror symmetries of the altermagnetic  nanostructure resulting in the spin-Hall effect that is present in Fig. \ref{spinHalldis} (b). However, since there is no mirror symmetry even in the ideal antiferromagnetic twin structure, the spin-Hall effect is typically stronger for the disordered antiferromagnetic dot ( Fig. \ref{spinHalldis} (a)) than for the disordered altermagnetic dot ( Fig. \ref{spinHalldis} (b)). Also, the disorder results in the inequality between the spin-Hall conductances for the two voltage leads shown in red and black in  Fig. \ref{spinHalldis}.

\subsubsection{Spin Filtering }

In the two-terminal case where leads 3 and 4 are removed from the disordered antiferromagnetic and altermagnetic quantum dots in Fig.\ref{dis} the disorder breaks the $\mathcal{IT}$ symmetry of the antiferromagnetic dot in Fig.\ref{dis} (b). The disordered altermagnetic dot in Fig.\ref{dis} (a) also does not have $\mathcal{IT}$ symmetry. Thus  according to the reasoning in Sec.\ref{filter}, in the two-terminal geometry the disordered altermagnetic dot in Fig.\ref{dis} (a) and the disordered antiferromagnetic dot in Fig.\ref{dis} (b) should both exhibit spin filtering qualitatively similar to that in Fig.\ref{spinFltr}. This has been confirmed by the present numerical work.

\subsection{Symmetry Breaking Leads}
\label{break}

If the configuration of the leads attached to the quantum dot breaks the C$_4$$\mathcal{T}$ symmetry for an altermagnetic dot or the $\mathcal{IT}$ symmetry for its antiferromagnetic twin, and also all mirror symmetries, then the present work predicts the transport properties of the dot to be qualitatively similar to those of dots with disorder discussed in Sec. \ref{disorderly} above. 

As an example, consider the Lieb  lattice quantum dots in Fig.\ref{lieb}. For the symmetric lead configuration shown in the figure, the altermagnetic dot exhibits no anomalous Hall effect and no spin-Hall effect, but does exhibit spin filtering. The antiferromagnetic dot exhibits an anomalous Hall effect and spin-Hall effect but no spin filtering. However, if lead 2 in Fig.\ref{lieb} is displaced to the right by even a single atomic site while the other leads are unchanged this breaks the C$_4$$\mathcal{T}$ symmetry, the $\mathcal{IT}$ symmetry and all mirror symmetries. As a result the altermagnetic dot acquires a giant anomalous Hall resistance similar to that of its antiferromagnetic twin for Fermi energies in the tunneling regime. At other values of the Fermi energy the altermagnetic dot acquires an anomalous Hall  resistance that is typically an order of magnitude weaker than that of its antiferromagnetic twin.
The broken symmetry also endows the altermagnetic dot with a spin-Hall effect comparable in overall strength to that of its antiferromagnetic twin.
With the broken lead symmetry the altermagnetic dot remains a spin filter and its antiferromagnetic twin becomes a spin filter of comparable efficiency.

This sensitivity of transport to a atomic scale displacements of leads is consistent with what is observed experimentally and understood theoretically in other nanoscale systems such as atomic and molecular wires bridging metal electrodes.\cite{Geo} 

 \section{Summary}
\label{S}

Altermagnets like antiferromagnets have no macroscopic magnetization, but unlike antiferromagnets they exhibit spin-split band structures.
The differences between the properties of altermagnetic and antiferromagnetic crystals have been attributed to differing symmetry operations  connecting the crystal sublattices of opposite spin in these materials.

This paper explores theoretically for the first time the transport properties of altermagnetic quantum dots with  C$_4$$\mathcal{T}$ symmetries and compares them with those of their antiferromagnetic twins with $\mathcal{IT}$ symmetries. For quantum dots that (together with conducting leads connected to them) obey these symmetries this work  makes the following predictions: 

1. The anomalous Hall effect, including giant anomalous Hall resistance in the tunneling regime, is predicted for antiferromagnetic dots but not for altermagnetic dots. 

2. The spin-Hall effect is predicted for both  antiferromagnetic and altermagnetic dots, excepts for those with mirror symmetries. 

3. Spin filtering (in two terminal arragements where the Hall voltage leads are decoupled from the dot but nothing else is changed) is predicted for altermagnetic dots but not for  antiferromagnetic dots.

These predictions have been made above for examples of quantum dot geometries based on the
1/5-depleted square lattice,  the square-octagon lattice and  the Lieb lattice
with arm-chair, zigzag, square or octagon edges and various lead arrangements, all conforming to C$_4$$\mathcal{T}$ or $\mathcal{IT}$
symmetries.

If the relevant C$_4$$\mathcal{T}$ or $\mathcal{IT}$ symmetries and all mirror symmetries are broken by disorder in the dot and/or by the arrangement chosen for the leads then the present work predicts both the altermagnetic and  antiferromagnetic dot to exhibit the anomalous Hall effect, including giant anomalous Hall resistance in the tunneling regime, the spin-Hall effect and two terminal spin filtering. I.e., symmetry breaking blurs the qualitative differences between the transport properties of altermagnetic and  antiferromagnetic quantum dots.

\begin{acknowledgments}
This research was supported by the Digital Research Alliance of Canada.
\end{acknowledgments}

%% ----------------------------------------------------------------------------------------------------
{

\end{document}